\documentclass[preprint,aps,tightenlines,floatfix,showpacs]{revtex4}
\usepackage{graphics}
\usepackage{bm}
\usepackage{amsmath}
\usepackage{amssymb}

\begin{document}

\title{
\bf Coupled-channel calculation of bound and resonant spectra of
$^9_\Lambda $Be  and $^{13}_\Lambda $C hypernuclei. } %$
%-------------------------------------- Authors --------------------
\author{ L. Canton$^{(1)}$} %$
\email{luciano.canton@pd.infn.it}
\author{ K. Amos$^{(2)}$}  %$
\email{amos@physics.unimelb.edu.au}
\author{ S. Karataglidis$^{(3)}$} %$
\email{kara@physics.unimelb.edu.au}
\author{ J. P. Svenne$^{(4)}$} %$
\email{svenne@physics.umanitoba.ca}

\affiliation{
$^{(1)}$Istituto Nazionale di Fisica Nucleare, sezione di Padova,\\
via Marzolo 8, Padova I-35131, Italia}
\affiliation{
$^{(2)}$School of Physics, University of Melbourne, Victoria 3010,
Australia}
\affiliation{
$^{(3)}$Department of Physics and Electronics, Rhodes University,
Grahamstown 6140, South Africa}
\affiliation{
$^{(4)}$Department of Physics and Astronomy, University of Manitoba,
and Winnipeg Institute for Theoretical Physics, Winnipeg, Manitoba,
Canada R3T 2N2}

\date{December 6, 2009}


%%%%%%%%%%%%%%%%%% ABSTRACT
\begin{abstract}

A Multi-Channel Algebraic Scattering (MCAS) approach has been used  to
analyze the spectra of two hypernuclear systems,  $^9_\Lambda $Be  and
$^{13}_\Lambda $C. The splitting of the two odd-parity  excited levels
($\frac{1}{2}^-$  and  $\frac{3}{2}^-$)  at  11  MeV  excitation    in
$^{13}_\Lambda $C  is  driven  mainly  by  the  weak $\Lambda$-nucleus
spin-orbit force,   but the  splittings  of  the  $\frac{3}{2}^+$  and
$\frac{5}{2}^+$ levels in both $^{9}_\Lambda $Be and $^{13}_\Lambda $C
have a  different  origin.       These cases appear to be dominated by
coupling to the collective $2^+$ states of the core nuclei.      Using
simple phenomenological potentials as input to the MCAS method,    the
observed   splitting   and   level   ordering   in $^{9}_\Lambda$Be is
reproduced with the addition of a  weak  spin-spin  interaction acting
between the hyperon and the spin of the excited target.   With no such
spin-spin  interaction,   the  level  ordering  in $^{9}_\Lambda$Be is
inverted with respect to that currently observed. In both hypernuclei,
our calculations suggest that there are additional low-lying  resonant
states   in   the $\Lambda$-nucleus continua.
\end{abstract}

\maketitle

\section{Introduction}
\label{sec1}

The study of hypernuclei opens up a new dimension for probing 
the structure and reactions of nuclei, with the presence of 
the strangeness degree of freedom. One can carry out nuclear 
physics with a ``tagged'' fermion which is distinguishable from the 
others resulting in various exotic hyperon-effects, and consider new 
systems such as excess-neutron systems with an hyperon tag. 

Hypernuclei have been the subject of quite extensive experimental
studies and their excitation spectra are now accessible
with high-quality measurements involving reactions such as
($\pi^+$, $K^+$), ($K^-$, $\pi^-$) and ($e,e'K^+
$)~\cite{pi91,ha96,aj98,ho01,ta94,ag05,ag05b,mi03,ga05,io07},
especially with the new generation
of germanium-based detectors for high-precision $\gamma$-ray
spectroscopy. The advances in $\Lambda$-hypernuclear
spectroscopy have been reviewed recently~\cite{Ha06,Ta08}.
In those reviews, important aspects of the role of the
hyperon in nuclear systems have been highlighted.
Effects such as the ``shrinkage'' of the
hypernuclear size~\cite{ta01},
the (weak) role of spin-orbit splitting in light hypernuclei,
and the connection of the hypernuclear excitation spectra
with the in-medium hyperon-nucleon interaction were noted;
the last with regard to its specific spin structure in particular.

Most theoretical studies of hypernuclei structure have sought
to use experimental data to define properties of the
underlying $\Lambda$-nucleon interaction, with the many-nucleon
properties specified by shell-model states~\cite{mi01,mi05}
or by mean-field theories~\cite{vr98}.
Recently, a mean-field model~\cite{ma08}
assessed the level of collectivity in hypernuclei
through particle-hole RPA-like calculations, with the indication
that the role of compound collectivity in hypernuclei
is much less important than in ordinary nuclei. While this
gives support to mean-field approaches in the description
of hypernuclear spectra, further exploration of
the role of ordinary nuclear collectivity due to the core
dynamics coupled to the single-hyperon motion is warranted.

Additionally, and specifically for the two hypernuclear systems,
$^9_\Lambda $Be and $^{13}_\Lambda $C, that we
consider, cluster-model calculations have been made~\cite{Hi00}.
In this context,
the two hypernuclei were considered as a three-body
($\alpha\alpha\Lambda$) and a four-body ($\alpha\alpha\alpha\Lambda$) system,
respectively. That study sought to
define characteristics of the $\alpha\Lambda$ interaction
and, by de-convolution, of the $\Lambda$-nucleon one. Of
particular concern was the $\Lambda$-nucleon spin-orbit term.
With $^9_\Lambda $Be, Hiyama {\it et al.}~\cite{Hi00} took the ground
state to be an $s_{\frac{1}{2}}$-wave $\Lambda$ coupled to the $0^+$ ground
state of $^8$Be. They then evaluated energies of a doublet ($\frac{3}{2}^+,
\frac{5}{2}^+$) considered dominantly composed of an
$s_{\frac{1}{2}}$-wave $\Lambda$
coupled to the $2^+_1$ state of $^8$Be. Splittings of between
0.08 and 0.2 MeV were found using diverse parameter sets.   From
their cluster model calculation of $^9_\Lambda $Be,  Filikhin, Gal, and
Suslov~\cite{Fi04}  found that the $p$-wave $\Lambda$-$\alpha$
interaction played a role in defining the binding energy of
$^9_\Lambda $Be as well as in determining the
excitation energies of the positive-parity doublet.
They suggest that the residual mismatch to experimental values may
be  attributed to  three-body forces.

In this paper we explore the possible description of the levels
of both hypernuclei $^9_\Lambda $Be and $^{13}_\Lambda $C
in terms of a phenomenological $\Lambda$-light mass
nucleus interaction which explicitly couples the hyperon to the collective
low-lying states of the ordinary nuclear core.  The phenomenological 
character of an appropriate $\Lambda$-light mass nucleus interaction was  
established in recent reviews of hypernuclear theory and
experiments~\cite{Mi07,Ha06,Ta08}. Essentially
it has a central depth of $\sim 30$ MeV (with a  Woods-Saxon form)
and a spin-orbit attribute considerably weaker than that
of a nucleon-nucleus interaction. 

In this work we do not investigate the microscopic origin of such 
model interaction. That would require a folding made to deduce
$\Lambda$-nucleus (optical) potentials from some specific 
$\Lambda$-nucleon effective force model. This would be similar to the 
nucleon-nucleus optical-potential model constructed by 
$g$-folding~\cite{Am00}, which combines detailed nuclear structure 
information with $NN$ effective interactions that have central, 
two-body spin-orbit, and tensor components,
with the results being a very non-local, energy and medium dependent
optical potential that has central and spin-orbit character.

We employ the MCAS approach to particle-nucleus structure and
scattering~\cite{Am03} since it emphasises the couplings of
single-particle dynamics  with  low-lying collective excitations 
of the ordinary nuclear core.  
The MCAS approach is holistic in that it considers the
system of a hyperon and a nucleus as a dynamic coupling of that 
hyperon to the whole nucleus, allowing complete freedom to the 
channel-coupling dynamics of the selected states. 
That this dynamics may play  a significant role
in defining hypernuclear states has been conjectured recently by
Hashimoto and Tamura~\cite{Ha06}, who noted that when a hyperon is added 
to a nucleus, nuclear properties of the compound vary from those of 
that when an extra nucleon is considered in place of the hyperon. 
A quote from that article on this aspect reads :``Nuclear properties such
as shape, size, symmetry, cluster and shell structures, and collective 
motions may be altered".

The role of MCAS studies to date, has been to analyze bound and resonant
spectra to support and interpret experimental investigations.
When the effects of the Pauli principle were incorporated in nucleon-nucleus
dynamics, the method has been  shown to describe, consistently,
the bound and resonant spectra of normal (zero-strangeness) light-mass
nuclei~\cite{Am03,Pi05,Ca06,Ca06a}.  Starting with the
properties of spectra of non-strange nuclei, we  consider the
modifications to the Hamiltonians in MCAS that are   required  to
describe hyperon-nucleus dynamics.  In particular, we analyze low-lying
level structures of two  $p$-shell $\Lambda$ hypernuclei with regards to
the structure of the hypernuclear doublet levels.  We consider splittings
that have been measured recently and, as well,  find level structures
that  are just above the $\Lambda$-nucleus scattering threshold.
Perhaps, they may be observed in future experiments.

We discuss, in brief, the MCAS method in the next section, while the
results for $^9_\Lambda{} $Be and $^{13}_\Lambda $C are presented and
discussed in Sections~\ref{Be-results} and \ref{C-results} respectively.
Conclusions are given in Section~\ref{conclusions}.

\section{The MCAS approach.}

%%%%%%%%%%% Multichannel scatt theory %%%%%%%%%%%%%%%%%%%%%%%%%%%%%%%%%%%
\label{multiT}

We describe the interaction Hamiltonian for the hyperon-core system
within a phenomenological potential model which couples
ground state and low-lying core excitations
%%%%%%%%%%%%%%%%%%%%%%%%%%%%%%%%%%%%%%%%%%%%%%%%%%%%%%%%%%%%%%%%%%%%
\begin{equation}
V_{cc'}(r)\ = \sum_{n}\ V_n\ {\cal O}_n
\ f_n(r,R,\theta_{\bf r, R})\, .
\end{equation}
%%%%%%%%%%%%%%%%%%%%%%%%%%%%%%%%%%%%%%%%%%%%%%%%%%%%%%%%%%%%%%%%%%%%%%
We assume that these low-lying excitations are
of rotational collective nature generated by a permanent
quadrupole deformation of the core subsystem.
The deformed nuclear surface is then described
by the radial angular dependence, $R=R_0\,[1+\beta_2P_2(\theta)]$,
and leads to a second-order expansion (in the $\beta_2$ parameter)
of the nuclear-interaction  functions $f_n$
%%%%%%%%%%%%%%%%%%%%%%%%%%%%%%%%%%%%%%%%%%%%%%%%%%%%%%%%%%%%%%%%%%%%%%%%%
\begin{equation}
f_n(r,R,\theta) =
f^{(0)}_n(r) - \beta_2R_0P_2(\theta)\frac{d}{dr}f^{(0)}_n(r)
  +
\frac{\beta_2^2R_0^2}{2\sqrt{\pi}}
\left( P_0-\frac{2\sqrt{5}}{7}P_2(\theta)+\frac{2}{7}P_4(\theta)\right)
\frac{d^2}{dr^2}f^{(0)}_n(r)\ .
\end{equation}
%%%%%%%%%%%%%%%%%%%%%%%%%%%%%%%%%%%%%%%%%%%%%%%%%%%%%%%%%%%%%%%%

In the calculations we have made, phenomenological  Woods-Saxon
functions and their derivatives have been used.  
Also we have used just three types of operators, ${\cal O}_n$, 
namely central $n = 0$, spin-orbit, $n = \ell s$,  and spin-spin, 
$n = s I$. We have not considered other possible terms
in the hyperon-nucleus interaction since the amount
of data available  for hypernuclei is too limited as yet to allow for the
determination of other small components. For this and other details
in the general MCAS scheme we refer to Ref.~\cite{Am03}.
The potential we employ generates a coupled-channel description
of the dynamics, which couples the ground-state to the
first $2^+$ state, as well as to possible higher excitations
of the core such as the second-order $0^+$, $2^+$, and $4^+$ levels.

In the MCAS approach,   sturmian functions
are chosen  as  a  basis  with  which   coupled-channel  interaction
potentials can be expanded.    Each element of the interaction matrix then
becomes a sum of separable interactions.    The analytic properties of the
$S$-matrix from a separable   Schr\"{o}dinger  potential in momentum space
allow a  full algebraic solution of the  multichannel  scattering problem.
With MCAS then, one starts
with a coupled-channel system that describes  nucleon-core dynamics   with
explicit inclusion of the low-lying excitations of the core. The
$\Lambda$-nucleus system is analogous as the $\Lambda$ has spin $\frac{1}{2}$
and a similar mass to the nucleon.  However, the Pauli principle, which is
so important when dealing with the nucleon-nucleus systems~\cite{Ca05},
is not relevant in an MCAS treatment of the $\Lambda$-nucleus ones.

With the MCAS method,
for each allowed scattering spin-parity,  $J^\pi$,
one solves Lippmann-Schwinger integral equations   in momentum space,
i.e.
%%%%%%%%%%%%%%%%%%%%%%%%%%%%%%%%%%%%%%%%%%%%%%%%%%%%%%%%%%%%%%%%%%%%%%%%%%
\begin{align}
T_{cc'}(p,q;E) = V_{cc'}(p,q)\ +\ \frac{2\mu}{\hbar^2}
&\left[ \sum_{c'' = 1}^{\rm open} \int_0^\infty V_{cc''}(p,x)\
\frac{x^2}{k^2_{c''} - x^2 + i\epsilon}\ T_{c''c'}(x,q;E)
\ dx \right.
\nonumber\\
& \left.- \sum_{c'' = 1}^{\rm closed} \int_0^\infty
V_{cc''}(p,x)\ \frac{x^2}{h^2_{c''} + x^2} \
T_{c''c'}(x,q;E) \ dx \right]\ ,
\label{multiTeq}
\end{align}
%%%%%%%%%%%%%%%%%%%%%%%%%%%%%%%%%%%%%%%%%%%%%%%%%%%%%%%%%%%%%%%%%%%%
where the index  $c$ denotes the set of quantum numbers that identify each
channel uniquely.  To use this approach, one must specify input  potential
matrices $V_{cc'}^{J^\pi}(p,q)$.              The open and closed channels
contributions have channel-wave numbers $k_c$ and $h_c$ which
are defined  for
$E > {\cal E}_c$ and $E < {\cal E}_c$ respectively. ${\cal E}_c$
are the excitation energies of the associated target state in channel $c$,
and $\mu$  is the reduced
mass.  Solutions of Eq.~(\ref{multiTeq}) are found using expansions of the
potential matrix elements in (finite) sums of energy-independent separable
terms,
%%%%%%%%%%%%%%%%%%%%%%%%%%%%%%%%%%%%%%%%%%%%%%%%%%%%%%%%%%%%%%%%%%
\begin{equation}
V_{cc'}(p,q) \sim  \sum^N_{n = 1} \chi_{cn}(p)\
\eta^{-1}_n\ \chi_{c'n}(q)\ .
\label{finiteS}
\end{equation}
%%%%%%%%%%%%%%%%%%%%%%%%%%%%%%%%%%%%%%%%%%%%%%%%%%%%%%%%%%%%%%%%%%%%%%
The method involves expansion form factors, $\chi_{cn}(q)$, that are built
using sturmian functions associated with the actual (coordinate space) model
interaction potential matrices, $V_{cc'}(r)$, we take to    describe the
coupled-channel problem. The eigenvalues of the sturmian equations
for all channels are the set of $\eta_n$.
When they are listed in order of decreasing
size, each term in the expansion, Eq.~(\ref{finiteS}), scales similarly
permitting truncation of the series. That makes the matrix problem one of
finite dimension.

The link between the multichannel $T$- and the scattering ($S$-) matrices
involves the numerical integration of a Green's function matrix,
%%%%%%%%%%%%%%%%%%%%%%%%%%%%%%%%%%%%%%%%%%%%%%%%%%%%%%%%%%%%%%%%%%%%%%%%%
\begin{equation}
\left( {G}_0 \right)_{nn'} =
\frac{2\mu}{\hbar^2}\left[ \sum_{c = 1}^{\rm open}
\int_0^\infty \chi_{cn}(x) \frac{x^2}{k_c^2 - x^2 + i\epsilon}
\chi_{cn'}(x)\ dx
  - \sum_{c = 1}^{\rm closed} \int_0^\infty
\chi_{cn}(x) \frac{x^2}{h_c^2 + x^2} \chi_{cn'}(x)\ dx
\right]\ ,
\label{xiGels}
\end{equation}
%%%%%%%%%%%%%%%%%%%%%%%%%%%%%%%%%%%%%%%%%%%%%%%%%%%%%%%%%%%%%%%%%%%%%%%%
with a diagonal eigenvalue matrix,
$\left( \eta \right)_{nn'} = \eta_n\ \delta_{nn'}$.
The bound states of the compound system are defined by the  zeros  of the
matrix determinant for energy $E < 0$.          They link to the zeros of
$\{ \left| \mbox{\boldmath $\eta$}-{\bf G}_0\right| \}$
when all channels in Eq.~(\ref{xiGels}) are closed.

Elastic scattering observables are determined by  the  on-shell properties
($k_1 = k_1^\prime = k$)    of the scattering matrices.    For the elastic
scattering  of  spin    $\frac{1}{2}$   particles  from spin zero targets,
$c=c'=1$,  and the $S$-matrix,
$S_{11} \equiv S_{\ell}^J = S_{\ell}^{(\pm)}$, is
%%%%%%%%%%%%%%%%%%%%%%%%%%%%%%%%%%%%%%%%%%%%%%%%%%%%%%%%%%%%%%%%%%%%%%%%%
\begin{equation}
S_{11} = 1 - i \pi \frac{2\mu}{\hbar^2}k \sum_{nn'=1}^M \ \chi_{1n}(k)
\frac{1}{\sqrt{\eta_n}} \left[\left({\bf 1} -
\mbox{\boldmath $\eta$}^{-\frac{1}{2}}
{\bf G}_0\mbox{\boldmath $\eta$}^{-\frac{1}{2}}
\right)^{-1}\right]_{nn'} \frac{1}{\sqrt{\eta_{n'}}}
\chi_{1n'}(k)\, .
\end{equation}
%%%%%%%%%%%%%%%%%%%%%%%%%%%%%%%%%%%%%%%%%%%%%%%%%%%%%%%%%%%%%%%%%%%%%
Diagonalizing the complex-symmetric matrix,
%%%%%%%%%%%%%%%%%%%%%%%%%%%%%%%%%%%%%%%%%%%%%%%%%%%%%%%%%%%%%%%%%%%%%%%%
\begin{equation}
\sum_{n'=1}^N
{\eta_n}^{-\frac{1}{2}}\left[{\bf G}_0\right]_{nn'}
{\eta_{n'}}^{-\frac{1}{2}} \tilde{Q}_{n'r} = \zeta_r \tilde{Q}_{nr} \, ,
\label{eigen}
\end{equation}
%%%%%%%%%%%%%%%%%%%%%%%%%%%%%%%%%%%%%%%%%%%%%%%%%%%%%%%%%%%%%%%%%%%%%%%%%%
establishes  the  evolution  of  the complex eigenvalues    $\zeta_r$ with
respect to energy, and that defines the  resonance attributes.
It can be shown~\cite{Am03} that
%%%%%%%%%%%%%%%%%%%%%%%%%%%%%%%%%%%%%%%%%%%%%%%%%%%%%%%%%%%%%%%%%%%%%%%%%%
\begin{equation}
\left[\left({\bf 1} -  \mbox{\boldmath $\eta$}^{-\frac{1}{2}}
{\bf G}_0\mbox{\boldmath $\eta$}^{-\frac{1}{2}} \right)^{-1}\right]_{nn'}
=\sum_{r=1}^N \tilde{Q}_{nr}\frac{1}{1-\zeta_r}\tilde{Q}_{n'r} \, .
\end{equation}
%%%%%%%%%%%%%%%%%%%%%%%%%%%%%%%%%%%%%%%%%%%%%%%%%%%%%%%%%%%%%%%%%%%%%%%%%%
Resonant
behaviour occurs when one of the complex $\zeta_r$ eigenvalues passes close
to the point (1,0) in the Gauss plane, since
the elastic
channel $S$-matrix has a pole structure at the corresponding energy.

\section{
The $^{9}_\Lambda $Be system}
\label{Be-results}

Before considering the $^{9}_\Lambda $Be hypernucleus,  we consider
first the ordinary (non-strange) $^{9}$Be nucleus.
To use the MCAS scheme, we view $^{9}$Be as an extended two-body system.   We
presume the neutron-$^8$Be core system involves coupling to the first  and
second states of $^8$Be  (a 2$_1^+$ at 3.03 MeV and a 4$^+$ at  11.35 MeV,
respectively). Though those two states are resonances in their own  right,
in calculation we take them to be discrete and to couple  via a rotational
model prescription as has been used heretofore~\cite{Am03,Pi05,Ca06}.

In this study we consider only the positive-parity states
of the compound nucleus, and employ a most simple
Hamiltonian interaction. A more extensive analysis of this system
for states of both parities has been made recently~\cite{Fr08}.
The effect of the
alpha-decay process of the two excited levels of $^{8}$Be
was included in the coupled-channel dynamics.
While we refer to
that work for full details, it is sufficient here to note that
the decay-widths of the two excited $^8$Be states had an important
effect on the widths of the compound levels, but did not change
significantly their energy centroids.

The neutron-core coupled-channel Hamiltonian we use for the
positive-parity states is  defined  by  the parameter values
listed under the heading $n$+${}^8$Be in Table~\ref{table3}.
%%%%%%%%%%%%%%%%%%%%%%%%%%%%%%%%%%%%%%%%%%%%%%%%%%%%%%%%%%%%%%%%%%%%%%
\begin{table}[ht]
\begin{ruledtabular}
\caption
{\label{table3} Parameter values of the $n$-${}^8$Be interaction. \\ }

\begin{tabular}{cccc}
$V_0$ (MeV) & $-$43.2 & $R_0$ (fm) & 2.7 \\
$V_{\ell s}$ (MeV) & 10.0 & $a$ (fm) & 0.65\\
  &  & $\beta_2$ & 0.7  \\
\end{tabular}
\end{ruledtabular}
\end{table}
%%%%%%%%%%%%%%%%%%%%%%%%%%%%%%%%%%%%%%%%%%%%%%%%%%%%%%%%%%%%%%%%%%%%%%%%%%%%%%%
We do not present extensive  aspects since
we seek only the bulk features of the $n$-$^8$Be system
to construct a phenomenologically plausible
interaction for the $\Lambda$-$^8$Be system. 
%We remind that only a
%few levels of $^9_\Lambda $Be have been observed up to date, and they have
%all positive parity.

Since the deeply bound $s$ states are
already occupied by the core nucleons, an appropriate OPP term has been
included in the nucleon-core Hamiltonian~\cite{Ca05}.
However, when considering the hyperon-nucleus system,
these OPP terms are not used since the hyperon is
distinguishable from the nucleons and access to the
deeply-bound orbits is not hindered by Pauli exclusion.
Some ${}^9$Be states that result from the MCAS evaluation are listed in
Table~\ref{table4}.
%%%%%%%%%%%%%%%%%%%%%%%%%%%%%%%%%%%%%%%%%%%%%%%%%%%%%%%%%%%%%%%%%%%%%
\begin{table}[ht]
\begin{ruledtabular}
\caption{
The low lying positive-parity spectra of $^{9}$Be 
compared with the results of an MCAS evaluation of $n$+${}^8$Be (in MeV).
\label{table4}\\}
\begin{tabular}{ccc}
$J^\pi$  & Exp.~\cite{Ti04} & Theory \\
\hline
$\frac{9}{2}^+$   &  5.0946  &   4.82  \\
$\frac{3}{2}^+$   &  3.0386   &   3.02 \\
$\frac{5}{2}^+$   &  1.3836  &   1.06  \\
$\frac{1}{2}^+$   &  0.0186  &   -0.05  \\
\end{tabular}
\end{ruledtabular}
\end{table}
%%%%%%%%%%%%%%%%%%%%%%%%%%%%%%%%%%%%%%%%%%%%%%%%%%%%%%%%%%%%%%%%%%%%%%
Clearly the bulk features of the positive-parity spectrum
can be reproduced with a  remarkably simple coupled-channel Hamiltonian.

Most of the recent work on $^9_\Lambda$ Be concentrates on the 
ground level and on the splitting of the two first excited levels
$\frac{5}{2}^+$ and $\frac{3}{2}^+$ \cite{Mi07,Ta05a,Ta08, Ak02}.
These states are all of positive parity and therefore 
herein we also consider only the  positive-parity spectrum.
However, we should mention for completeness 
that earlier theoretical works extensively 
dealt with predictions on low-energy negative-parity states.
Such states were first predicted by Dalitz and Gal, using 
shell model \cite{Da76, Au83}. Amongst these, specific hypernuclear states 
termed ``supersymmetric states'' have been considered and later described as 
``genuine hypernuclear states'' in Refs.~\cite{Mo83, Ya88}, with the $\Lambda$ 
hyperon moving on a $p$-wave orbit parallel to the symmetric $\alpha-\alpha$ 
axis. Based on comparison with the predicted spectra, a few peaks at about 
6 and 10 MeV excitation in the E336 experiment~\cite{Ha98}
were interpreted as generated by those specific states.

To initiate study of the $\Lambda$-$^8$Be  system we applied
scalings to the $n$-$^8$Be Hamiltonian adjusting the  parameter values by
reducing the central potential by 30\%,   reducing the interaction radius
by 15\%, and reducing the spin-orbit strength by an order of magnitude.
Small adjustments were then made to give
best results. The final values of the  coupled-channel potential
parameters are given in Table~\ref{table3Lambda}. In this Table, 
we consider two sets of parameters, one with the onset of a small
spin-spin component of the interaction (`Case 2').
%%%%%%%%%%%%%%%%%%%%%%%%%%%%%%%%%%%%%%%%%%%%%%%%%%%%%%%%%%%%%%%%%%%%%%%%%%%%
\begin{table}[ht]
\begin{ruledtabular}
\caption
{\label{table3Lambda} Strengths of the $\Lambda$-${}^{8}$Be interaction
with $R_0 = 2.3$ fm., $a = 0.65$ fm., and $\beta_2 = 0.7$}
\begin{tabular}{ccc}
  & {Case 1}  & {Case 2} \\
\hline
$V_0$ (MeV) & $-$26.4 & $-$26.4\\
$V_{\ell s}$ (MeV) & 0.35 & 0.35 \\
$V_{sI}$ (MeV) & 0.0 & $-$0.1 \\
\end{tabular}
\end{ruledtabular}
\end{table}
%%%%%%%%%%%%%%%%%%%%%%%%%%%%%%%%%%%%%%%%%%%%%%%%%%%%%%%%%%%%%%%%%%%%%%%%%%
 Our choice of the spin-spin interaction strength was made 
 to be comparable with that of the spin-orbit interaction and by 
 the knowledge that the hypernuclear state splittings are themselves
 small.  There are small spin-spin, spin-orbit, and tensor interactions in 
 most phenomenological $\Lambda$-nucleon interactions (see table 6
 in Ref.~\cite{mi01} for an example of their effects). Such components
 albeit stronger, are present in nucleon-nucleon forces, the  
 folding of which with microscopic models of
 structure~\cite{Am00} lead to nucleon-nucleus central and spin-orbit
 interactions. So one should allow for such terms in  effective
 $\Lambda$-nucleus potentials.  Furthermore, and in parallel with the 
 phenomenological interactions required in MCAS studies of nucleon-nucleus 
 systems~\cite{Am03}-\cite{Fr08},
 we expect there to be ($\Lambda$-nucleus) spin-spin terms. 
 Given the small strengths of both spin-orbit and spin-spin
 interactions in the $\Lambda$-nucleon potentials, we anticipate
 the both will also be small in the $\Lambda$-nucleus interactions. 
 Herein our purpose is to identify the effects of such an
 interaction rather than ascertaining what would be an optimum value.
 For the study of the $\Lambda-{}^8$Be system, values of 0 and -0.1 for 
 this term suffice.  We do study effects of variation in this term
 with our study of the $\Lambda-{}^{12}$C system, though only over
 the range between $\pm 0.1$.

In Table~\ref{Table1-BeLam}, we give the spectrum calculated
with the MCAS approach and the two sets of parameters given in  
Table~\ref{table3Lambda}. We report the experimental 
binding energy of $^9_\Lambda $Be as determined from emulsion data~\cite{Ju73},
but have also indicated in brackets the additional, newer value
obtained by the E336 experiment at KEK (see Ref.~\cite{Ha06}).
It has been already observed by Hashimoto and Tamura~\cite{Ha06}
that, while the emulsion data for $^7_\Lambda $Li and
$^{13}_\Lambda $C agreed well with the KEK-experiment results,
there was significant disagreement between those 
regarding the $^9_\Lambda $Be ground-state binding energy, with  
the reason  for this disagreement not known.   
Also, we consider for comparison shell-model results.
These results \cite{mi05} were obtained by considering not only the $\Lambda N$ and
$\Sigma N$ interactions but also the $\Lambda \Sigma$ coupling, which was shown to make an
important contribution to the spacing of the $1^+$ and $0^+$ states in $^4_\Lambda$H 
and $^4_\Lambda$He. The $\Lambda$ was assumed to be in the $0s$ orbit while the nucleons were
assumed to be in the $0p$-shell. Comparison with the bound-state spectrum obtained from MCAS
is quite good.

%Hashimoto and Tamura~\cite{Ha06} observed
%that, while the emulsion data for $^7_\Lambda $Li and
%$^{13}_\Lambda $C agreed with the KEK-experiment results,
%there was significant disagreement between those
%regarding the $^9_\Lambda $Be ground
%state binding energy.  The reason  for this disagreement is not known.
%%%%%%%%%%%%%%%%%%%%%%%%%%%%%%%%%%%%%%%%%%%%%%%%%%%%%%%%%%%%%%%%%%%%%%
%\begin{table}[ht]
%\begin{ruledtabular}
%\caption{
%The positive-parity spectrum, in MeV, of ${}^9_\Lambda $Be.
%The columns labelled ``Case 1'' and ``Case 2''
%refer, respectively,  to calculations made without and with the
%spin-spin term in the potential.  The spin-spin strength, $V_{s\cdot I}$,
%was $-$0.1 MeV.  The widths of resonant states are given in brackets.
%\label{Table1-BeLam}}
%\begin{tabular}{cccc}
%$J^\pi$  & Exp.~\cite{Ak02} & Case 1 & Case 2 \\
%\hline
%$\frac{7}{2}^+$   &  $--$   &     4.791 (4.1 keV) &  4.92 (4.9 keV)      \\
%$\frac{9}{2}^+$   &  $--$   &     4.788  (4.4 keV) & 4.68 (3.8 keV)     \\
%$\frac{3}{2}^+$   &  $-$3.64  &   $-$3.70  &   $-$3.63           \\
%$\frac{5}{2}^+$   &  $-$3.69  &   $-$3.65  &   $-$3.70           \\
%$\frac{1}{2}^+$   &  $-$6.71 ($-$5.99$^\dagger$) &   $-$6.73 &    $-$6.73    \\
%\end{tabular}
%\end{ruledtabular}
%\vskip 0.4cm
%{\small
%$^\dagger$ This value result was found from a 1998 experiment~\cite{Ha06}
%(see the discussion in the text).}
%\end{table}
%%%%%%%%%%%%%%%%%%%%%%%%%%%%%%%%%%%%%%%%%%%%%%%%%%%%%%%%%%%%%%%%%%%%%%%
\begin{table}[ht]
\begin{ruledtabular}
\caption{
The positive-parity spectrum, in MeV, of ${}^9_\Lambda $Be. 
The columns labelled `Case 1' and `Case 2' 
refer, respectively,  to calculations made without and with the 
spin-spin term in the potential.  The spin-spin strength, $V_{s\cdot I}$,
was $-$0.1 MeV.  The widths of resonant states are given in brackets. Comparison is
also made with the results of a shell model calculation \cite{mi05}, where the ground state binding
energy has been set to the measured value for comparison.
\label{Table1-BeLam}}
\begin{tabular}{ccccc}
$J^\pi$  & Exp.~\cite{Ak02} & Case 1 & Case 2 & Shell Model~\cite{mi05} \\
\hline
$\frac{7}{2}^+$   &  $--$   &     4.791 (4.1 keV) &  4.92 (4.9 keV)  &    \\
$\frac{9}{2}^+$   &  $--$   &     4.788  (4.4 keV) & 4.68 (3.8 keV)   &  \\
$\frac{3}{2}^+$   &  $-$3.64  &   $-$3.70  &   $-$3.63  &  $-$3.66        \\
$\frac{5}{2}^+$   &  $-$3.69  &   $-$3.65  &   $-$3.70   & $-$3.71        \\
$\frac{1}{2}^+$   &  $-$6.71 ($-$5.99$^\dagger$) &   $-$6.73 &    $-$6.73   & $-$6.71  \\
\end{tabular}
\end{ruledtabular}
\vskip 0.4cm
{\small
$^\dagger$ This value result was found from a 1998 experiment~\cite{Ha06}
(see the discussion in the text).}
\end{table}
%%%%%%%%%%%%%%%%%%%%%%%%%%%%%%%%%%%%%%%%%%%%%%%%%%%%%%%%%%%%%%%%%%%%%%

It is interesting to observe that we get the correct size of
this fine splitting between these two states with a simple phenomenological
model consisting only of a central and a spin-orbit potential
(`Case 1' in Table~\ref{table3Lambda}). Indeed, 
assuming no $sI$ coupling,
the  magnitude of the  splitting between the $\frac{3}{2}^+$ and
$\frac{5}{2}^+$ states is very small, but 
consistent with the separation value recently measured~\cite{Ta05}. 
But it is predicted that the $\frac{3}{2}^+$ state to be the lower, at
variance with the recent analysis~\cite{Ta05a}, where 
the $\frac{3}{2}^+$ state was assessed to be the less bound of the pair.

    The shell model calculations~\cite{mi05} gave the observed
  splitting and order of the three bound states of ${}^9_\Lambda$Be
  and from them it was noted that the $\Lambda$-nucleon spin-orbit
  interaction was primarily responsible for the 
  $\frac{3}{2}^+$-$\frac{5}{2}^+$ states splitting. Other components
  in the chosen interaction (spin-spin, tensor, etc.) contributed
  values that essentially cancelled. But, it is to be noted that
  the $\Lambda$-nucleon force is phenomenological. So it is the parameter 
  set chosen for this force that gives that splitting and state order 
  obtained with the shell-model calculation.

\begin{figure}[h]
\scalebox{0.8}{\includegraphics*{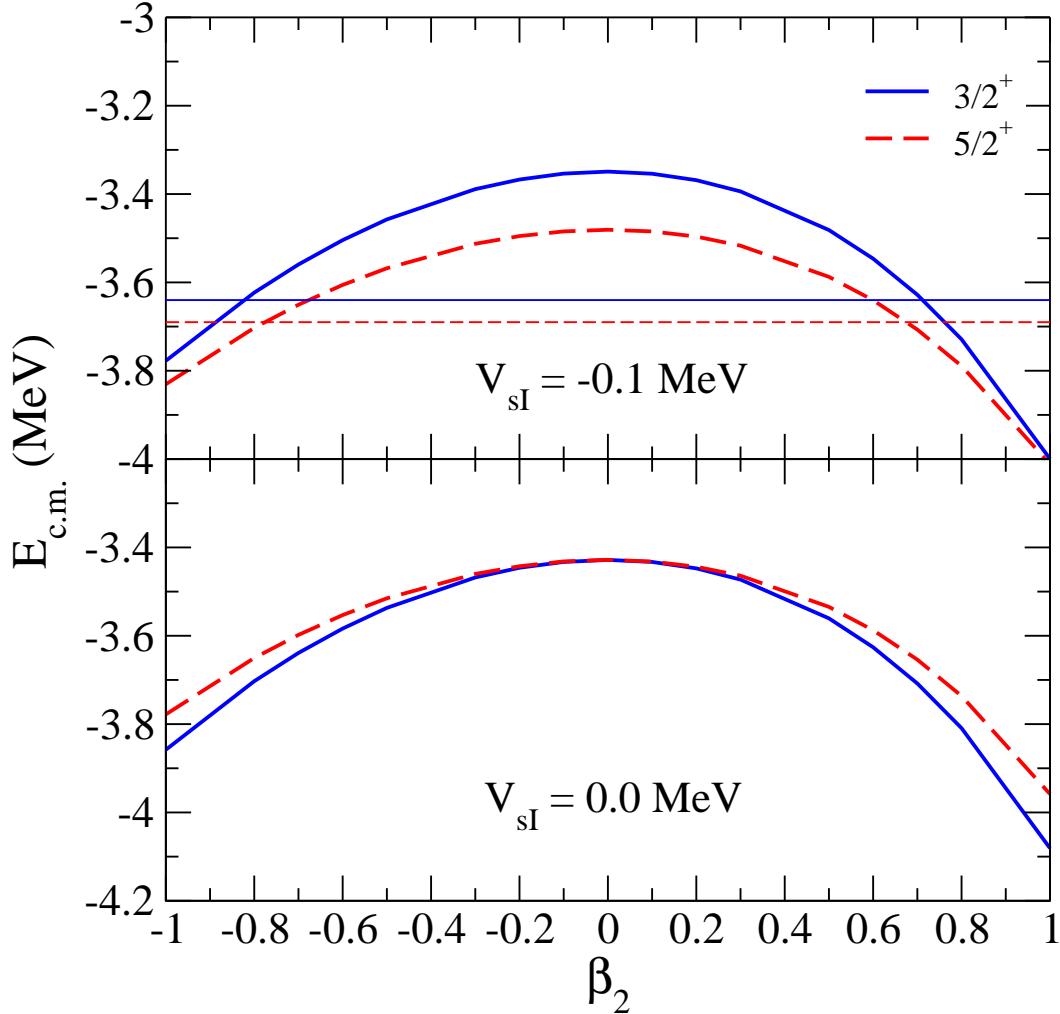}}
\caption{\label{Canton-fig1}
The binding energies of the $J^\pi = \frac{3}{2}^+$ (solid curves) and
$J^\pi = \frac{5}{2}^+$ (dashed curves) states in $^9_\Lambda$Be
with variation of the deformation parameter. The horizontal lines
denote the experimental values of the two states.}
\end{figure}

To achieve the correct level ordering with MCAS, requires the introduction of a small
spin-spin contribution in the $\Lambda$-$^8$Be phenomenological
interaction (`Case 2'). This need is emphasised by considering the variation
with deformation of the binding energies of the two positive-parity states in question.
This is shown in Fig.~\ref{Canton-fig1}
%%%%%%%%%%%%%%%%%%%%%%%%%%%%%%%%%%%%%%%%%%%%%%%%%%%%%%%%%%%%%%%
%%%%%%%%%%%%%%%%%%%%%%%%%%%%%%%%%%%%%%%%%%%%%%%%%%%%%%%%%%%%%%%%%%%%%
for the two parameter sets, `Case 1' (bottom panel) and `Case 2'
(top panel). With no spin-spin interaction strength (Case 1), the
splitting of the states is small but with the $\frac{3}{2}^+$ state
always the more bound of the pair. In contrast are the results when the
spin-spin interaction strength is finite though small as shown in
the top panel of the figure. Now the $\frac{5}{2}^+$ state is always
the more bound of the pair. The actual size of the splitting varies
slightly with the deformation but more so with the value of $V_{sI}$.

At 4.7 MeV above the scattering threshold, we predict two additional
positive-parity states (resonances).
These are formed by  coupling the
4$^+$ state in ${}^8$Be with a $\Lambda $ in the $s_\frac{1}{2}$ state.
The widths of these resonances shown in brackets in
Table~\ref{Table1-BeLam}
were calculated assuming that the
${}^8$Be-core 4$^+$ state has zero width. If we consider
the $\alpha$-decay probability of this $4^+$ level,
then the widths reported in the table can be expected to increase quite
significantly~\cite{Fr08}.

\section{
The $^{13}_\Lambda $C system}
\label{C-results}

       The bound and scattering properties of the $^{13}$C (and ${}^{13}$N)
  systems have been
studied extensively using the MCAS approach~\cite{Am03,Pi05,Sv06}.
Good reproduction of the
low-lying spectra, sub-threshold and resonant states,   and of the elastic
scattering cross-section and polarization data have been obtained with a
relatively simple model Hamiltonian.  But such only resulted
if the incoming nucleon (on ${}^{12}$C) was blocked from access to 
the phase-space already
occupied by closed shells (in this case,        the $s_{\frac{1}{2}}$ and
$p_{\frac{3}{2}}$ levels).       With MCAS, that is achieved by using the
Orthogonalizing Pseudo-Potential (OPP) method~\cite{Ca05,Ca07}.

There is no such requirement in using MCAS to study the $\Lambda$-${}^{12}$C 
system.
However,   the  depth of the  $\Lambda$-nucleus optical potential is about
$\frac{2}{3}$ that of the standard nucleon-nucleus one.     The spin-orbit
strength also is much smaller, an order of magnitude smaller, than that of
the corresponding nucleon-nucleus system.
Additionally, the potential radius is $\sim 15$\%
smaller than used for the $n$+${}^{12}$C system.
Starting from that set of parameter values,
with but small adjustements, the coupled-channel potential interaction 
used in MCAS describes the known spectrum.  The resultant final
potential parameters we have used are listed in Table~\ref{table1}.
%%%%%%%%%%%%%%%%%%%%%%%%%%%%%%%%%%%%%%%%%%%%%%%%%%%%%%%%%%%%%%%%%%%%%%%%%%%%
\begin{table}[ht]
\begin{ruledtabular}
\caption
{\label{table1} Strengths of the $\Lambda$-${}^{12}$C interaction
with $R_0 = 2.6$ fm., $a = 0.6$ fm., and $\beta_2 = $-$0.52$ }
\begin{tabular}{ccccc}
  &\multicolumn{2}{c}{Case 1}  &\multicolumn{2}{c}{Case 2} \\
  & $\pi = -1$ & $\pi = +1$ & $\pi = -1$ & $\pi = +1$\\
\hline
$V_0$ (MeV) & $-$28.9 & $-$30.4 & $-$28.9 & $-$30.4\\
$V_{\ell s}$ (MeV) & 0.35 & 0.35 & 0.35 & 0.35 \\
$V_{sI}$ (MeV) & 0.0 & 0.0 & $-$0.1 & $-$0.1 \\
\end{tabular}
\end{ruledtabular}
\end{table}
%%%%%%%%%%%%%%%%%%%%%%%%%%%%%%%%%%%%%%%%%%%%%%%%%%%%%%%%%%%%%%%%%%%%%%%%%%
Again, the parameters identified as `Case 1' were those we found with
the spin-spin interaction strength set to zero,
while those defined as `Case 2' had
the small spin-spin interaction strength listed.
The parities of the channel interactions are designated by $\pi = \pm 1$.

%%%%%%%%%%%%%%%%%%%%%%%%%%%%%%%%%%%%%%%%%%%%%%%%%%%%%%%%%%%%%%%%%%%%%
%\begin{table}[ht]
%\begin{ruledtabular}
%\caption{Spectra of
%$^{13}_\Lambda$C
%\label{table2}}
%\begin{tabular}{ccll}
%$J^\pi$  & Exp.~\cite{Ko02} & Case 1 & Case 2\\
%\hline
%$\frac{1}{2}^-$   &  $---$    &   +4.65  (0.21 MeV)  &   +4.66 (0.23 MeV) \\
%$\frac{3}{2}^-$   &  $---$    &   +4.64  (0.22 MeV)  &   +4.63 (0.21 MeV) \\
%$\frac{5}{2}^-$   &  $---$    &   +4.28  (1.0 keV)   &   +4.31  (1.0 keV) \\
%$\frac{7}{2}^-$   &  $---$    &   +4.17  (1.0 keV)   &   +4.14  (1.0 keV) \\
%$\frac{3}{2}^-$   &  $---$    &   +3.10  (0.1 keV)   &   +3.15  (0.1 keV) \\
%$\frac{5}{2}^-$   &  $---$    & +3.05 ($>$0.1 keV)   & +3.02 ($>$0.1 keV) \\
%$\frac{1}{2}^-$   &  $-$0.708 &    $-$0.74\          &   $-$0.74\         \\
%$\frac{3}{2}^-$   &  $-$0.86\ &    $-$0.89\          &   $-$0.89\         \\
%$\frac{1}{2}^+$   &  $---$    &    $-$4.12\          &   $-$4.12\         \\
%$\frac{3}{2}^+$   &  $-$6.81\ &    $-$7.177          &   $-$7.08\         \\
%$\frac{5}{2}^+$   &  $-$6.81\ &    $-$7.178          &   $-$7.24\         \\
%$\frac{1}{2}^+$   &  $-$11.69 &   $-$11.68           &   $-$11.68         \\
%\end{tabular}
%\end{ruledtabular}
%\end{table}
%%%%%%%%%%%%%%%%%%%%%%%%%%%%%%%%%%%%%%%%%%%%%%%%%%%%%%%%%%%%%%%%%%%%%%%%%
%%%%%%%%%%%%%%%%%%%%%%%%%%%%%%%%%%%%%%%%%%%%%%%%%%%%%%%%%%%%%%%%%%%%%
\begin{table}[ht]
\begin{ruledtabular}
\caption{Spectra of
$^{13}_\Lambda$C with energies in MeV. Nomenclature is as for Table.~\ref{Table1-BeLam}. 
The shell model results are those received from Millener~\cite{Mi09}.
\label{table2}}
\begin{tabular}{ccllc}
$J^\pi$  & Exp.~\cite{Ko02} & Case 1 & Case 2 & Shell Model~\cite{Mi09} \\
\hline
$\frac{1}{2}^-$   &  $---$    &   +4.65  (0.21 MeV)  &   +4.66 (0.23 MeV) & \\
$\frac{3}{2}^-$   &  $---$    &   +4.64  (0.22 MeV)  &   +4.63 (0.21 MeV) & \\
$\frac{5}{2}^-$   &  $---$    &   +4.28  (1.0 keV)   &   +4.31  (1.0 keV) & \\
$\frac{7}{2}^-$   &  $---$    &   +4.17  (1.0 keV)   &   +4.14  (1.0 keV) & \\
$\frac{3}{2}^-$   &  $---$    &   +3.10  (0.1 keV)   &   +3.15  (0.1 keV) & \\
$\frac{5}{2}^-$   &  $---$    & +3.05 ($>$0.1 keV)   & +3.02 ($>$0.1 keV) & \\
$\frac{1}{2}^-$   &  $-$0.708 &    $-$0.74\          &   $-$0.74\      &   \\
$\frac{3}{2}^-$   &  $-$0.86\ &    $-$0.89\          &   $-$0.89\      &   \\
$\frac{1}{2}^+$   &  $---$    &    $-$4.12\          &   $-$4.12\       &  \\
$\frac{3}{2}^+$   &  $-$6.81\ &    $-$7.177          &   $-$7.08\   &   $-$6.22   \\
$\frac{5}{2}^+$   &  $-$6.81\ &    $-$7.178          &   $-$7.24\   &  $-$6.19    \\
$\frac{1}{2}^+$   &  $-$11.69 &   $-$11.68           &   $-$11.68  & $-$10.95        \\
\end{tabular}
\end{ruledtabular}
\end{table}
%%%%%%%%%%%%%%%%%%%%%%%%%%%%%%%%%%%%%%%%%%%%%%%%%%%%%%%%%%%%%%%%%%%%%%%%%

To date, four state energies  have been measured and they are
listed in Table~\ref{table2}
in  the  column  labelled  `Exp'.  A splitting  between  the
$\frac{3}{2}^+$ and $\frac{5}{2}^+$ states is not resolved as yet.   The
theoretical spectra, however, contain a richer  structure as shown by
the results listed under the `Case 1' and `Case 2' columns in the table.
Both model calculations predict a $\frac{1}{2}^+$ bound state  at 4.12
MeV below threshold. As the spin-spin interaction has no effect upon its
excitation energy, this state corresponds to an
$s_{\frac{1}{2}}$-$\Lambda$ coupled to the $0_2^+$ state at 7.65 MeV
in $^{12}$C, which is an highly exotic state as it corresponds to the 
coupling of the hyperon to the superdeformed Hoyle state. This state is not
predicted by the shell model \cite{Mi09}, as the $0p$-shell model of the underlying
structure of $^{12}$C cannot predict the Hoyle state. The ground state and low-lying spectrum from
the shell model for $^{13}_\Lambda$C, however, agree generally well with the predictions from MCAS.
The $\frac{1}{2}^+$ state we expect at 7.56 MeV excitation in
$^{13}_\Lambda $C has been placed much higher in excitation (12.2 MeV)
in the cluster model evaluations~\cite{Hi00}. That is due to the strong
state dependence of the  $\Lambda$-nucleus interaction, found by
$s$-wave folding in that model~\cite{Ya85}.
We have still to wait for more detailed experimental investigations of that spectrum
to decide whether it is a strong state dependence effect or simply
the extremely weak natural excitation of the second $0^+$ from the
ground state of $^{12}$C that explains the position of a second
$\frac{1}{2}^+$ state in the spectrum of this hypernucleus.
Note that a similar
state has been predicted by MCAS calculations of $n$+${}^{12}$C to be
in the spectrum of the non-strange $^{13}$C.
That state has spin-parity $J^\pi=\frac{1}{2}^-$, lies 2.68 MeV
above the scattering threshold,  and though unobserved, is partner to
a mirror state~\cite{Pi05} of such structure  in $^{13}$N at
6.97 MeV above the $p$-${}^{12}$C threshold. That is
1.3 MeV higher than where a $J^\pi=\frac{1}{2}^-$ state has been
observed in $^{13}$N.

Additionally,
a set of six odd-parity states of $^{13}_\Lambda $C are predicted to be
just a  few MeV above the scattering threshold.
They are states formed by coupling of a $p_{\frac{3}{2}}$- and of a
$p_{\frac{1}{2}}$-$\Lambda$ with the $2^+_1$ state in ${}^{12}$C.
Without deformation (and no spin-spin interaction) they would form
a degenerate
quartet ($\frac{1}{2}^-, \frac{3}{2}^-, \frac{5}{2}^-$, and  $\frac{7}{2}^-,$)
and a degenerate doublet ($\frac{3}{2}^-$ and  $\frac{5}{2}^-$) of states.
Deformation and spin-spin effects then break those degeneracies and
mix states of like spin-parity.  Since these six states are
embedded in the $\Lambda$-$^{12}$C continuum, they are resonances
and their widths are listed (in brackets) in Table~\ref{table2}.
These resonance states are very narrow save
for the doublet of $\frac{3}{2}^-\vert_3$ and $\frac{1}{2}^-\vert_2$
resonances.  We have also calculated  the
corresponding excitation function, assuming a low-energy $\Lambda$-$^{12}$C
elastic scattering process.

%%%%%%%%%%%%%%%%%%%%%%%%%%%%%%%%%%%%%%%%%%%%%%%%%%%%%%%%%%%%%%%
\begin{figure}[h]
\scalebox{0.9}{\includegraphics*{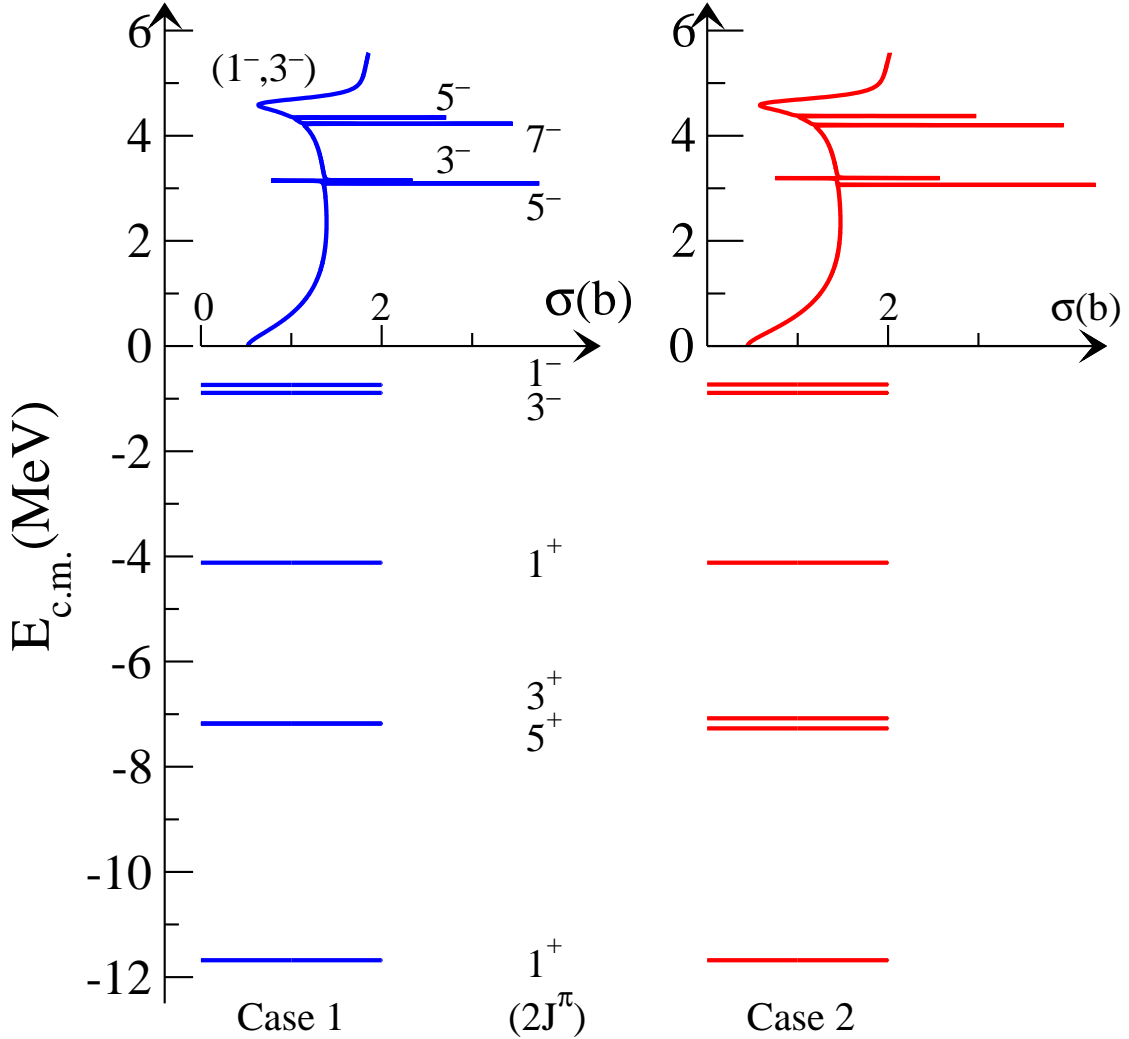}}
\caption{\label{Canton-fig2}
The spectra and total elastic cross sections for
$\Lambda$-$^{12}$C scattering resulting from MCAS calculations
made using two sets of parameter values.}
\end{figure}
%%%%%%%%%%%%%%%%%%%%%%%%%%%%%%%%%%%%%%%%%%%%%%%%%%%%%%%%%%%%%%%%%%%%%
The spectra for $^{13}_\Lambda $C are depicted in
Fig.~\ref{Canton-fig2}. Below the $\Lambda-^{12}$C threshold
the discrete bound states are shown while above that threshold the
theoretical cross sections are given for both the case of $V_{sI}$ =
0.0 and $-$0.1 MeV.
Of prime interest is the fine splitting of the bound  $\frac{1}{2}^-$  and
$\frac{3}{2}^-$ levels,   observed respectively at 10.98 MeV and 10.83 MeV
above the ground state~\cite{Ta05}.    We note that there is also a very
small splitting expected between two bound levels,     $\frac{5}{2}^+$ and
$\frac{3}{2}^+$, at the excitation energy of $\sim$4.88 MeV. Our exploratory
calculation suggests a splitting of $\sim$160 keV  for both doublets.
Experimentally, the splitting of the $\frac{1}{2}^-$ and
$\frac{3}{2}^-$ states was found to be 152 $\pm$ 54 $\pm$ 36 keV.      The
splitting of the more bound positive-parity states has not been determined
quantitatively to date.

\begin{figure}
\scalebox{0.7}{\includegraphics{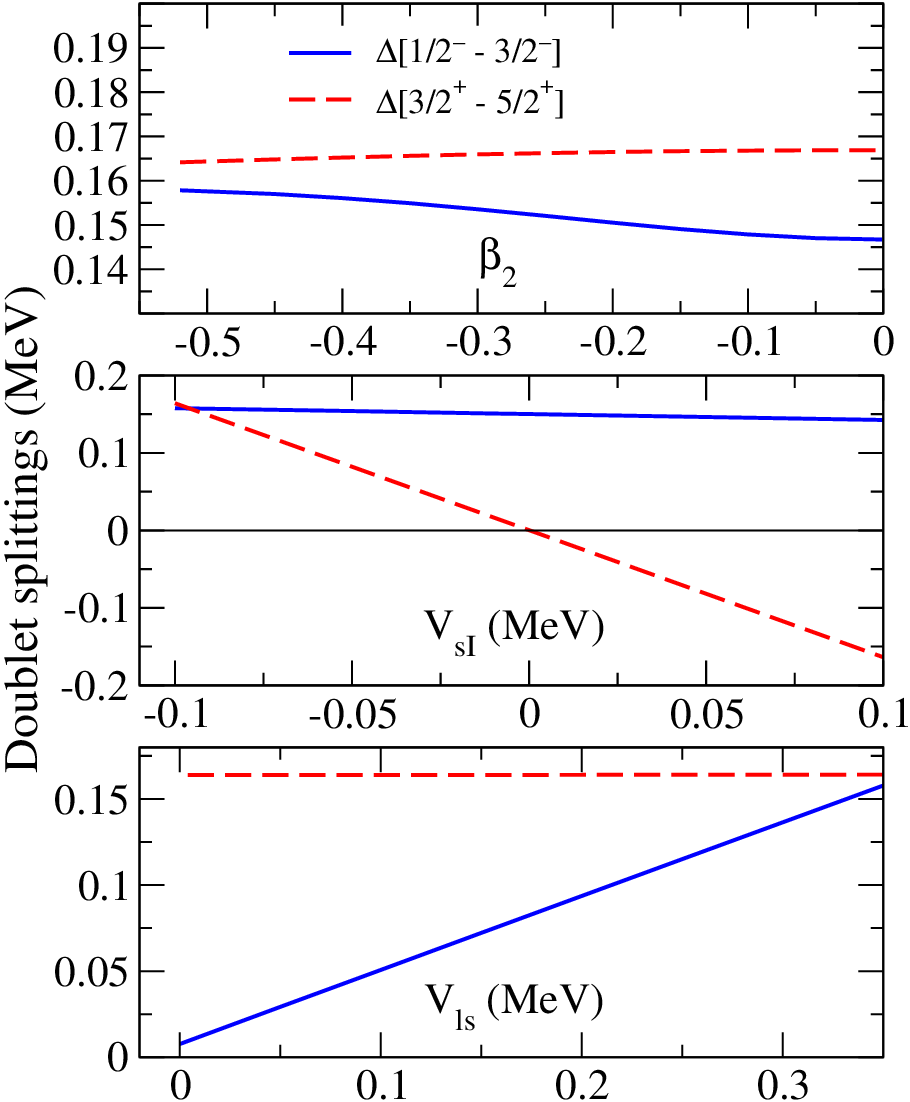}}
\caption{ \label{Canton-fig3}
MCAS values of the two level splittings in
$^{13}_\Lambda $C with respect to changes in the $^{12}$C
quadrupole-deformation parameter $\beta_2$, in the
spin-spin hyperon-${}^{12}$C coupling interaction, and in the
spin-orbit hyperon-${}^{12}$C coupling interaction.}
\end{figure}

The results of an analysis of parameter   variation on these splittings is
presented in Fig.~\ref{Canton-fig3}.
In this figure with basic parameter values of $\beta_2 = -0.52$,
$V_{sI} = -0.1$ MeV, and $V_{\ell s} = 0.35$ MeV, variation of the deformation
parameter (with the other two parameters held fixed) gave the results
depicted in the top panel. Likewise with the other two parameter values
fixed, variations of the state energies with the strengths of the
spin-spin and of the spin-orbit interactions are shown in the middle
and bottom panels, respectively.  We first notice that both splittings have
only a weak dependence on the $\beta_2$ parameter value, however the
actual binding energies of the $\frac{1}{2}^-$ and $\frac{3}{2}^-$ states
vary by 600 keV over the range of $\beta_2$ chosen.
The splitting of the two odd-parity states has essentially no
  dependence upon the spin-spin
($V_{sI}$) coupling, thus indicating that origin of the splitting of those
negative-parity states is  almost solely the result of spin-orbit effects.
But  the  splitting  of  the  positive-parity  ($\frac{5}{2}^+$        and
$\frac{3}{2}^+$) levels is dominated by the $V_{sI}$ coupling that links
with the excitation of the $2_1^+$ (4.43 MeV) state in $^{12}$C.
%----------------------------------------- Figure --------------------
%----------------------------------------- Figure --------------------
Indeed, as shown in the middle panel of Fig.~\ref{Canton-fig3}, there is
no splitting of the positive-parity doublet for $V_{sI} = 0$, with a
linear trend for small non-zero values.  But the $\frac{5}{2}^+$ state
is more bound than the $\frac{3}{2}^+$ one for negative values of $V_{sI}$.
Over the range of $V_{sI}$ strength shown, the splitting of the
negative-parity states changes little, while that of
the positive-parity states changes by $300$ keV.
Finally we show the effect of the spin-orbit interaction terms in the
bottom panel. With variation of this strength, the positive-parity
doublet splitting is essentially unchanged, as are the individual state
binding energies. This is expected since the states involve coupling
of an $s_{\frac{1}{2}}$-$\Lambda$ particle. On the other hand,
the splitting of the $\frac{1}{2}^-$-$\frac{3}{2}^-$ varies
linearly with  the strength of the spin-orbit force in the 
   $\Lambda$-${}^{12}$C interaction. These two states have also been
  found in shell model studies, taking them to be $p_{\frac{1}{2}}$
  and $p_{\frac{3}{2}}$ $\Lambda$ single-particle states coupled to the
  $0^+$ ground state of ${}^{12}$C. Their separation was thought to
  be a measure of the $\Lambda$-nucleon spin-orbit force. 
  With that prescription, spin-spin and tensor components of the
  two-body force were found to be important. However, 
  Millener~\cite{mi01,Mi09} 
  notes that these states will have components involving coupling to
  the $2^+_1$ state in ${}^{12}$C, that the ground state of ${}^{12}$C 
  has components  that bring into play other attributes in the
  interactions with the hyperon, and that the $p$-wave $\Lambda$
  orbits are weakly bound.

These properties indicate that one can  extract the $\Lambda$-nucleus
spin-orbit   strength   from   the   splitting  of the $\frac{1}{2}^-$ and
$\frac{3}{2}^-$ states. Using  the  experimental  information  of  a 152
keV splitting for the two
odd-parity levels, we settle upon a $\Lambda$-nucleus spin-orbit strength
of 0.35 MeV.      We also conclude that the current knowledge of experimental
spectra is insufficient  to  assess any importance of the
$\Lambda$-nucleus $V_{sI}$ coupling in $^{13}_\Lambda $C.
For this reason, for the
$\frac{5}{2}^+$, $\frac{3}{2}^+$  level  splitting we  considered the
${}^9_\Lambda $Be   system  for  which  accurate experimental 
information  on   the
splitting of such levels has been obtained~\cite{Ha06}.
That information leads to a significant
constraint on the $V_{sI}$ coupling.

\section{Summary and Conclusions}
\label{conclusions}

  The MCAS approach to define scattering of two clusters and 
 to describe the structure of the compound system formed of them 
 requires a matrix of coupled-channel interactions.
 These describe the pairwise interactions between the clusters, 
 in this case the hyperon and the nucleus. This interaction matrix is 
 related to the one we have found best describes the scattering 
 and compound system of a neutron and that same nucleus. 
 This approach to study hypernuclear spectroscopy is quite new 
 and reveals effects of coupled channel dynamics 
 in the spectroscopy which, to our knowledge, have not been 
 considered before.  As we consider what might be the 
 hyperon-nucleus coupled-channel interaction within collective
 model prescriptions of the nucleus, of course there may be 
 specifics that can be improved.  But the results offer an 
 explanation of aspects concerning the limited amount of assured 
 levels for these systems alternate to the usual 
 adjusted-interaction approaches.

Specifically, we have applied the MCAS approach to study
the excitation spectra of light hypernuclei
$^9_\Lambda $Be and $^{13}_\Lambda $C.
The theoretical approach emphasises the single-particle motion of the
hyperon in the mean-field of the ordinary nuclear core, which is however
coupled to its own low-lying collective motions.

The phenomenological hyperon-core potential was constructed
from the ordinary nucleon-core potential by removing Pauli-blocking
effects, making a 10-15\% shrinkage of the radius, using
a $\frac{2}{3}$ reduction of the strength of the central potential,
and reducing drastically the spin-orbit potential.
However, the coupling of the single hyperon
with the collective motion of the core is
rather important in defining the hypernuclear spectra.
The deformation parameter is essentially that required
in the associated nucleon-nucleus dynamics.

In light of the recently observed fine splitting of the excited
hypernuclear spectra for these two nuclei we conclude
that the  $\frac{1}{2}^-$-$\frac{3}{2}^-$ splitting in $^{13}_\Lambda $C
is dominated by the $\Lambda$-core spin-orbit interaction.
This splitting is largely insensitive to other factors such as the
deformation of the core or other spin-dependent components of the
$\Lambda$-core potential.  On the other hand, the fine splittings of the
$\frac{3}{2}^+$-$\frac{5}{2}^+$ doublet in
$^9_\Lambda $Be and $^{13}_\Lambda $C have a different dynamical origin.
They originate from the coupling of the $s_\frac{1}{2}$-$\Lambda$ single-particle motion
with the $2^+$ collective excitation of the core.
As it involves an $s$-wave $\Lambda$-particle,
it is independent of the spin-orbit interaction. While this splitting
varies slightly with the deformation of the core, it is
most sensitive to
the coupling of the $s_\frac{1}{2} \Lambda$ single-particle motion
with the $2^+$ collective excitation of the core through the
$V_{sI}$ interaction.
The recently measured structure of the fine
($\frac{3}{2}^+$-$\frac{5}{2}^+$) splitting in $^9_\Lambda $Be
enabled us to determine the sign and strength of that spin-spin
interaction. With the same strength for the spin-spin component
taken for the  $^{13}_\Lambda $C system, a similar splitting
of the $\frac{3}{2}^+$-$\frac{5}{2}^+$ doublet results.

Finally, using the scattering properties embedded in the MCAS scheme,
with the interaction potential so determined by the measured
hypernuclear low-lying levels, we expect there to be resonant states
in both hypernuclei at low excitation energies. The centroid
energies and widths of those resonances we anticipate
to be rather more sensitive to the deformation in these systems.

\begin{flushleft}
\section*{Acknowledgements}
\end{flushleft}

L.C. acknowledges support from the Italian MIUR-PRIN Project
``Fisica Teorica del Nucleo e dei Sistemi a Pi\`u Corpi'' during the early stages of this project.
K.A. acknowledges support from the International Exchange program of the
Australian Academy of Science.
S.K. acknowledges support
from the National Research Foundation (South Africa). J.P.S.
acknowledges
support
from the Natural Sciences and Engineering Research Council (NSERC),
Canada.

\bibliography{Hyper-paper}

\end{document}